\documentclass[%
twocolumn, 
pre,
superscriptaddress,
 aps,
floatfix,
]{revtex4-2}
\usepackage{graphicx}
\usepackage{amssymb}
\usepackage{color}
\usepackage{amsmath}
\usepackage[table,xcdraw]{xcolor}
\usepackage{mathrsfs}
\usepackage[bookmarks=true,colorlinks,urlcolor=blue,linkcolor=blue,anchorcolor=blue,citecolor=blue,unicode]{hyperref}
\usepackage{bookmark}
\usepackage{bm}
\graphicspath{{Figures/}}

\begin{document}
\preprint{Phys. Rev. E}

\title{Temporal rich club phenomenon and its formation mechanisms} 

\author{Mu-Yao Li}
  \affiliation{School of Business, East China University of Science and Technology, Shanghai 200237, China} %
  \affiliation{Research Center for Econophysics, East China University of Science and Technology, Shanghai 200237, China} %

\author{Yin-Ting Zhang}%
  \affiliation{School of Business, East China University of Science and Technology, Shanghai 200237, China} %
  \affiliation{Research Center for Econophysics, East China University of Science and Technology, Shanghai 200237, China} %

\author{Wei-Xing Zhou}
  \email{wxzhou@ecust.edu.cn}
  \affiliation{School of Business, East China University of Science and Technology, Shanghai 200237, China} %
  \affiliation{Research Center for Econophysics, East China University of Science and Technology, Shanghai 200237, China} %
 \affiliation{School of Mathematics, East China University of Science and Technology, Shanghai 200237, China} %

\date{\today}

\begin{abstract}

The temporal rich club (TRC) phenomenon is widespread in real systems, forming a tight and continuous collection of the prominent nodes that control the system. However, there is still a lack of sufficient understanding of the mechanisms of TRC formation. Here we use the international N-nutrient trade network as an example of an in-depth identification, analysis, and modeling of its TRC phenomenon. The system exhibits a statistically significant TRC phenomenon, with eight economies forming the cornerstone club. Our analysis reveals that node degree is the most influential factor in TRC formation compared to other variables. The mathematical evolution models we constructed propose that the TRC in the N-nutrient trade network arises from the coexistence of degree-homophily and path-dependence mechanisms. By comprehending these mechanisms, we introduce a novel perspective on TRC formation. Although our analysis is limited to the international trade
system, the methodology can be extended to analyze the mechanisms
underlying TRC emergence in other systems.
\end{abstract}

\maketitle

\section{Introduction}

In many real-world networks, it's common to observe tightly interconnected communities formed by dominant nodes in the system \citep{Zhou-Mondragon-2004-IEEECommunLett,Wei-Song-Xiu-Zhao-2018-ApplGeogr,Kim-Min-2020-CompStructBiotechnolJ,Bullmore-Sporns-2012-NatRevNeurosci,Hu-Zhu-2009-PhysicaA}, known as the rich club phenomenon \citep{Colizza-Flammini-Serrano-Vespignani-2006-NatPhys}. These nodes exhibit not only strong connections in the static network structure but also stability and synchronization in the temporal network, referred to as the temporal rich club (TRC) phenomenon \citep{Pedreschi-Battaglia-Barrat-2022-NatPhys}. While the static rich club has been extensively studied and applied \citep{Leo-Fleury-AlvarezHamelin-Sarraute-Karsai-2016-JRSocInterface,Tang-Dong-Lian-Tang-2020-FuturGenerCompSyst,Chinazzi-Fagiolo-Reyes-Schiavo-2013-JEconDynControl,Xu-Zhang-Small-2010-PhysRevE}, the TRC is a relatively new field. Since the temporal network setup retains more information and potential communities \citep{Li-Cornelius-Liu-Wang-Barabasi-2017-Science, Presigny-Holme-Barrat-2021-PhysRevE}, not every well-connected rich club will evolve into a TRC occupying a crucial position in the system's evolution. Investigating the formation mechanism of TRC is certainly worthwhile.

The mechanisms responsible for the formation of TRCs differ across various systems. To comprehend these mechanisms, we need to examine two perspectives. Firstly, we must identify the nodes that form the tightest structure in the static network. Different systems exhibit different performances of static rich clubs \citep{Colizza-Flammini-Serrano-Vespignani-2006-NatPhys}, and different richness selections lead to varied results \citep{Opsahl-Colizza-Panzarasa-Ramasco-2008-PhysRevLett,Serrano-2008-PhysRevE}. These differences are attributed to the mechanisms of link formation in networks \citep{Barabasi-Albert-1999-Science,Barabasi-Jeong-Neda-Ravasz-Schubert-Vicsek-2002-PhysicaA,Papadopoulos-Kitsak-Serrano-Boguna-Krioukov-2012-Nature,Yuan-Alabdulkareem-Pentland-2018-NatCommun}. Secondly, we must analyze the evolution process of temporal networks. Systems evolve differently over time \citep{Petri-Expert-2014-PhysRevE}, such as social networks \citep{Gelardi-LeBail-Barrat-Claidiere-2021-ProcRSocB-BiolSci} and power grid networks \citep{Hartmann-Sugar-2021-SciRep}, where the former has high changeability, while the latter remains relatively stable. Different modes of evolution result in separate mechanisms for the formation of TRCs. Hence, it is necessary to focus on a single system initially to identify, analyze, and simulate its TRC phenomenon to understand the formation mechanisms of TRCs.

The N-nutrient trade is worth attention. N-nutrient is a crucial input for agriculture, impacting food security and sustainability \citep{Zhang-Davidson-Mauzerall-Searchinger-Dumas-Shen-2015-Nature,Foley-Ramankutty-Brauman-Cassidy-Gerber-Johnston-Mueller-O'Connell-Ray-West-Balzer-Bennett-Carpenter-Hill-Monfreda-Polasky-Rockstrom-Sheehan-Siebert-Tilman-Zaks-2011-Nature}, and international N-nutrient trade plays a vital role in the N cycle \citep{Galloway-Townsend-Erisman-Bekunda-Cai-Freney-Martinelli-Seitzinger-Sutton-2008-Science, Gu-Ju-Chang-Ge-Vitousek-2015-ProcNatlAcadSciUSA}, which is linked to ecosystem health \citep{Canfield-Glazer-Falkowski-2010-Science}, the greenhouse effect \citep{Cui-Zhang-Reis-Wang-Wang-He-Chen-vanGrinsven-Gu-2023-NatSustain}, and climate change \citep{Ren-Zhang-Reis-Wang-Jin-Xu-Gu-2023-NatFood}. Moreover, it exhibits an oligopoly among a few major players \citep{Hernandez-Torero-2013-AgricEcon}, resembling a rich club phenomenon. A detailed exploration of the formation of this alliance aligns perfectly with our investigation into the causes of TRCs.

As an international commodity trade, the analysis of the TRC mechanism in N-nutrient networks naturally draws insights from the literature on international trade. Explanations encompass the persistence of comparative advantage \citep{Kogut-Zander-1992-OS}, the emergence of dominant firms \citep{Teece-Pisano-Shuen-1997-Smj}, the spatial organization of economic activity \citep{Baldwin-RobertNicoud-2014-JIntEcon}, and the impact of policies \citep{Helpman-Krugman-1987}. The gravity model \citep{McCallum-1995-AmEconRev}, a frequently used framework in trade research, has depicted the influence of various policies and distances \citep{Anderson-vanWincoop-2003-AmEconRev,Helpman-Melitz-Rubinstein-2008-QJEcon}. Additionally, complex network models have been employed to analyze the mechanisms of trade frictions \citep{Chaney-2014-AmEconRev,Jun-Alshamsi-Gao-Hidalgo-2020-JEE}. Leveraging these existing theories enhances our understanding of the TRC generation mechanism, positioning the international N-nutrient trade network as a suitable subject for study.

The rest of the paper is organized as follows: Section~\ref{S:Methods and data} describes the database, the methodology of building networks, and the calculations of TRC. Section~\ref{S:result:A} reports the identification results of the TRC phenomenon in the international N-trient trade network. Section~\ref{S:result:B} and Section~\ref{S:result:C} analyse the localized TRC of different richness and different initial year to find the main factors that lead to TRC. Section~\ref{S:result:D} construct the theoretical models to explain the formation mechanisms of its TRC phenomenon. A summary discussion is conducted in Section~\ref{S:disscusion}.

\section{Methods and data}
\label{S:Methods and data}

\subsection{Data}

The trade data is sourced from the United Nations Commodity Trade Database (https://comtrade.un.org), providing detailed trade flow data between economies for 30 years (1991–2020). We calculate year-by-year inter-economy N-nutrient trade data based on the nutrient conversion concentrations in Table~\ref{Table:fertilizerNutrientturn}, following the method provided by the Food and Agriculture Organization of the United Nations (https://fao.org). The data is measured in tons. To handle repeated trade data, we preprocess it by using information reported by importing economies as the basis, supplemented by data from exporting economies \cite{Feenstra-Lipsey-Deng-Ma-Mo-2005-NBER}. Importing economies' data is considered more credible due to its direct connection to import tariffs.

\begin{table}[ht]
\caption{N-nutrient content ratios in related commodity.}
\centering
\setlength{\tabcolsep}{3.5mm}
\begin{tabular}{p{0.28\linewidth}<{\centering} p{0.25\linewidth}<{\centering} p{0.2\linewidth}<{\centering}}
    \hline
    Commodity & HS code & N-content  \\
    \hline
    Urea & 310210 & 46\% \\
    Ammonium sulphate & 310221 & 21\%   \\
    Ammonium nitrate & 310230 & 33.5\%  \\
    Calcium ammonium nitrate and other mixtures with calcium carbonate & 310240 & 26\% \\
    Sodium nitrate & 310250 & 16\%  \\
    Urea and ammonium nitrate solutions & 310280 & 32\%\\
    Ammonia, anhydrous & 281410 & 82\%  \\
    Other nitrogenous fertilizers, n.e.c. & 281420, 310260, 282710, 283410, 310229, 310290, 310270 & 20\%  \\
    NPK fertilizers & 310520 & 15\%  \\
    Diammonium phosphate & 310530 & 18\%  \\
    Monoammonium phosphate & 310540 & 11\%  \\
    Other NP compounds & 310551, 310559 & 20\%  \\
    Potassium nitrate & 283421 & 13\%  \\
    \hline
   \label{Table:fertilizerNutrientturn}
\end{tabular}
\end{table}

\subsection{Network construction}

The international N-nutrient trade network is
a complex system composed of nodes (trading economies), links (trade), and link weights (trade volume). Its 30-year evolution is captured by a temporal network denoted as $\mathscr{G}$, comprising 30 single-layer networks. For a single-layer network, it aggregates all trade relations of all economies in a given year $t$, depicted as $G(t)=\left(V(t), E(t)\right)$. Here, the set of nodes $V(t)=\left\{v_{it} \right\}$ represents all economies (denoted by $v_{it}$) involved in the N-nutrient trade in year $t$. The set of links $E(t) = \left\{e_{ijt}\right\}$ includes the trade relations (denoted by $e_{ijt}$) of N-nutrient trade exported from economy $v_i$ (or economy $i$ for simplicity) to economy $v_j$ (or economy $j$ for simplicity) in year $t$.

For networks, $e_{ijt}$ is
\begin{equation}
    e_{ijt}= \begin{cases}0, \ &{\mathrm{if}} \ w_{ijt} = 0,\\
    1, \ &{\mathrm{if}} \ w_{ijt} > 0,
    \end{cases}
    \label{Eq:TRC:e}
\end{equation}
where $w_{ijt}$ is the N-nutrient trade volume (unit in tons) exported from economy $i$ to economy $j$ in year $t$. 

Converging $t=1991, 1992, \cdots, 2020$, the overall temporal network is described as $\mathscr{G} = \left\{G(t)\right\}$, including the set of all temporal nodes $\mathscr{V} = \bigcup_t V(t)$ and the set of all temporal links $\mathscr{E} = \bigcup_t E(t)$. $\mathscr{G}(t_1, t_2)$ represents the sub temporal network from $t_1$ to $t_2$. Note that $\mathscr{G}(t_1, t_1)$ is equal to $G(t_1)$. A temporal link can be described as $(i, j, t, w)$, denoting an interaction from node $i$ to node $j$ at time $t$ with weight $w$.

\subsection{The temporal rich club}

The temporal rich club phenomenon, as defined by \citet{Pedreschi-Battaglia-Barrat-2022-NatPhys}, is characterized by the maximum density of links between nodes with a minimum required richness, while the links need remain stable for a specific duration. The emergence of this phenomenon indicates that well-connected nodes in a temporal network tend to form simultaneous and stable structures.

The definition of TRC originates from the classical rich club concept in static networks \citep{Colizza-Flammini-Serrano-Vespignani-2006-NatPhys}. For a static undirected network with node richness denoted as $r$, the rich club coefficient is defined as the density of the subgraph $G_{>r}$ that only contains nodes $V_{> r}$ with richness greater than $r$ and links $E_{> r}$ between them:
\begin{equation}
\phi(r)=\frac{2 \sharp \left[E_{> r}\right]}{\sharp \left[V_{> r}\right] (\sharp \left[V_{> r}\right]-1)},
\label{Eq:TRC:phi_undirected}
\end{equation}
$\sharp \left[ \bf{X} \right]$ is the number of members in set $\bf{X}$. For a directed network, the rich club coefficient is expressed as:
\begin{equation}
\phi(r)=\frac{\sharp \left[E_{> r}\right]}{\sharp \left[V_{> r}\right] (\sharp \left[V_{> r}\right]-1)}.
\label{Eq:TRC:phi_directed}
\end{equation}
Extended to temporal networks, we introduce localized TRC coefficients:
\begin{equation}
\epsilon(r, t, \Delta)=\frac{\sharp \left[ \bigcap_t^{t+\Delta-1}E_{> r}(t)\right]}{\sharp \left[\mathscr{V}_{> r}\right] (\sharp \left[\mathscr{V}_{> r}\right]-1)}.
\label{Eq:TRC:phi}
\end{equation}
Here, $\mathscr{V}_{>r}$ represents the set of nodes with specific richness. All nodes associated with the links in $\bigcap_t^{t+\Delta-1} E_{> r}(t)$ are involved in $\mathscr{V}_{>r}$. The intersection of $\Delta$ subsets of the continuous network from $t$ to $t+\Delta-1$ gives the set that remains stable during this period. The density of this stable part is the localized TRC coefficient $\epsilon(r, t, \Delta)$. When $\Delta=1$, $\epsilon(r, t, 1)$ corresponds to the static rich club coefficient $\phi(r)$ of the network $G(t)$. As $\Delta$ increases, $\epsilon$ analyzes whether the core of the system at time $t$ remains consistently well-connected from $t$ to $t+\Delta-1$.

Furthermore, when considering the temporal network as a whole, the maximum density obtained across all initial years $t$ represents the corresponding TRC under the $\Delta$ parameter. The TRC coefficient is defined as:
\begin{equation}
M(r, \Delta) = \max_{t} \epsilon(r, t, \Delta).
\label{Eq:TRC:epsilon:max}
\end{equation}
$M(r, \Delta)$ is designed to quantify several aspects: (1) whether the static rich club patterns correspond to a structure that existed at some instant, (2) how dense and stable such a structure is, or (3) whether the rich club is formed by links that appeared at unrelated times. An increasing $M(r, \Delta)$ with $r$ indicates that the richest nodes tend to be increasingly connected with each other in a simultaneous and stable manner for a duration of at least $\Delta$. This requirement is distinct from distinguishing stable and unstable hubs, as $M(r, \Delta)$ focuses on the links between hubs. The simultaneous association between hubs contributes the most to the TRC.

However, it's crucial to consider the random effects in network evolution. Only the structural evolution beyond these random effects reveals the real characteristics. For instance, in a highway network, the built highways often remain continuously connected, resulting in a high $\epsilon$. This is a property of the overall network rather than the rich club. Therefore, it becomes imperative to construct a null model for calculating $M_{\mathrm{rnd}}(r, \Delta)$ and compare it with the actual value $M(r, \Delta)$. In the computation of the null model, a reshuffling procedure function $P[w, t]$ \citep{Gauvin-Genois-Karsai-Kivela-Takaguchi-Valdano-Vestergaard-2022-SIAMRev} is employed to permute the timestamps $t$ of all temporal links while keeping the node indices $i$ and $j$ constant. This process generates a series of simple random networks that can help analyze whether intrinsic forces exist between hubs in the actual system.

For the comparison of $M(r, \Delta)$ and $M_{\mathrm{rnd}}(r, \Delta)$, a ratio is usually employed to assess the existence of the TRC phenomenon. It is considered to exist when $\hat{\mu}(r, \Delta) = \frac{M(r, \Delta)}{M_{\mathrm{rnd}}(r, \Delta)} > 1$. However, relying solely on this criterion may not be precise enough \citep{Jiang-Zhou-2008-NewJPhys}. In this study, we propose adding a statistical test as a basis for determining whether a TRC phenomenon exists. The null hypothesis is that $\mu(r, \Delta)$ is not greater than 1. The $p$-value is calculated as follows:
\begin{equation}
p = \frac{\sharp[\hat{\mu}(r, \Delta) \leq 1]}{n},
\label{Eq:TRC:trendp}
\end{equation}
where $\sharp[\hat{\mu}(r, \Delta) \leq 1]$ counts the number of $\hat{\mu}(r, \Delta)$ values that are not greater than 1. As $n \rightarrow \infty$, the estimated bootstrap $p$-value will tend toward the ideal bootstrap $p$-value. In our case, $n$ is set to 1000. The smaller the $p$-value, the stronger the evidence against the null hypothesis, favoring the alternative hypothesis that the presence of the TRC phenomenon is statistically significant. By adopting the conventional significance level of $\alpha = 1\%$, the TRC phenomenon is statistically significant if $p < 1\%$. For the portion that passes the statistical test, we use the value of $\mu(r, \Delta) = \frac{M(r, \Delta)}{\left\langle M_{\mathrm{rnd}}(r, \Delta)\right\rangle}$ as a characterization, with a larger $\mu(r, \Delta)$ indicating a stronger TRC phenomenon.

\section{Results}
\label{S:result}

\subsection{Identification of the TRC phenomenon}
\label{S:result:A}

\begin{figure*}[ht!!!]
    \centering
    \includegraphics[width=0.96\linewidth]{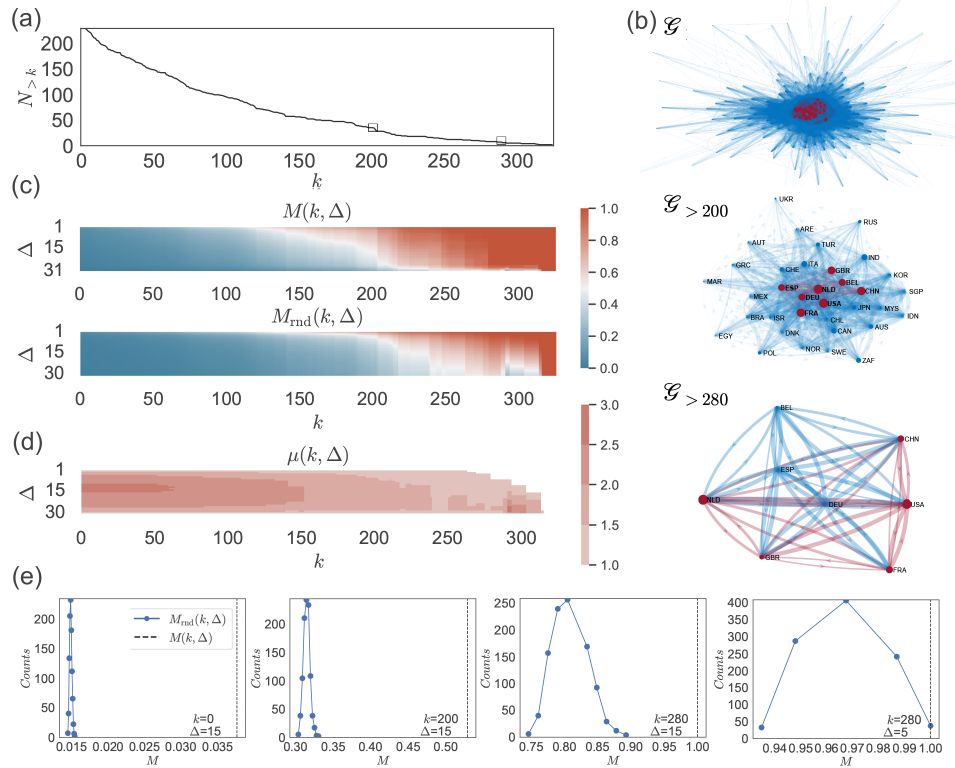}
    \caption{(a) The evolution of the number of nodes with an aggregate degree $> k$ as $k$ increases.
    (b) Upper: the overall network structure of $\mathscr{G}$, with the red (dark) area representing nodes and edges in $\mathscr{G}{> 200}$. Medium: the network structure of the sub-network $\mathscr{G}{> 200}$, with the red (dark) area indicating nodes and edges in $\mathscr{G}{> 280}$. Lower: the network structure of the sub-network $\mathscr{G}{> 280}$, with the red (dark) area representing nodes and edges in $\mathscr{G}_{> 300}$. The node size is proportional to the aggregate degree of the node. The width of the links is based on the number of recurrences. The layout of the three networks is calculated independently, considering only the existing structure.
    (c) Upper: TRC coefficients for the actual N-nutrient trade network $M(k, \Delta)$. Lower: TRC coefficients for the random networks generated by the null model $M_{\mathrm{rnd}}(k, \Delta)$.
    (d) The ratio $\mu(k, \Delta)$ of TRC coefficients for the actual network $M(k, \Delta)$ to TRC coefficients for the random networks $M_{\mathrm{rnd}}(k, \Delta)$. Those that do not pass the significance test are left blank.
    (e) The distribution of TRC coefficients in the random networks $M_{\mathrm{rnd}}(k, \Delta)$, denoted by the blue solid line, and the value of TRC coefficients in the actual network $M(k, \Delta)$, denoted by the black dashed line. The first three panels are statistically significant cases. The fourth panel depicts the statistically insignificant case.}
    \label{Fig:TRC:M}
\end{figure*}

For estimating the TRC phenomenon, the initial step involves determining which property of nodes represents richness. In many studies, the node's degree is the primary richness used \citep{Colizza-Flammini-Serrano-Vespignani-2006-NatPhys}. This choice is influenced by the Preferential Attachment (PA) model \citep{Barabasi-Albert-1999-Science, Albert-Barabasi-2000-PhysRevLett}, considered a reliable model for most realistic networks, where the probability of new links between nodes is proportional to their degree. Nodes with high degrees naturally form tight associations, and theoretical reference values have been provided $\phi_{unc}(k) \sim \frac{k^2}{\left\langle k \right\rangle N}$.
In temporal networks, degrees naturally extend to aggregate degrees $k$ \citep{Pedreschi-Battaglia-Barrat-2022-NatPhys}. Since the trade network is directed, the aggregate degree is defined as
\begin{equation}
k_i=k_i^{\mathrm{out}} + k_i^{\mathrm{in}}=\sum_{j} \left( E_{ij} + E_{ji} \right),
\label{Eq:TRC:K}
\end{equation}
where
\begin{equation}
E_{ij}= \begin{cases}1, \ &{\mathrm{if}} \ \sum_t e_{ijt} > 0\\
    0, \ & {\mathrm{otherwise}}
    \end{cases}
\end{equation}
Therefore, the aggregate degree $k$ is initially applied as the richness to identify whether a TRC phenomenon exists in the N-nutrient trade network.

Fig.~\ref{Fig:TRC:M}(a) reports the trend in the number of club members as the threshold increases. There are only two nodes in $\mathscr{G}_{>315}$ that remain permanently stable and fully connected: the Netherlands and the United States. This membership is so small as to be meaningless. The club $\mathscr{G}_{>280}$ can maintain a stable, fully connected state for more than 20 years, which is listed in the lower plot of Fig.~\ref{Fig:TRC:M}(b). Club members include the Netherlands, the United States, France, China, the United Kingdom, Belgium, Germany, and Spain. These eight economies constitute the core club of the N-nutrient trade network. As shown in the medium plot of Fig.\ref{Fig:TRC:M}(b), this club is central within a larger club $\mathscr{G}_{> 200}$ with 35 members. Furthermore, $\mathscr{G}_{>200}$ is the most central part of the overall network $\mathscr{G}$, as depicted in the upper plot of Fig.\ref{Fig:TRC:M}(b).

In Fig.~\ref{Fig:TRC:M}(c), we present the TRC coefficients from the N-nutrient trade network and the coefficients from the null model. $M(k,\Delta)$ exhibits a noticeable change at $k = 200$, occurring earlier than in the random network. The increasing TRC coefficient suggests a strengthening and stabilizing association between club members. For the club $\mathscr{G}{>200}$, the result of $M(200, 1)>0.8$ indicates a high density of immediate association, though not maintaining a stable state. The central club $\mathscr{G}{>280}$ boasts not only $M(280, 1)=1$ but also $M(280, 25)=1$. The long-term stability underscores the cornerstone role of this club for the system. Considering the TRC coefficients $M_{\mathrm{rnd}}(k, \Delta)$ of the random network generated by the null model, derived from the mean of 1000 random simulations, a crucial observation is that $M_{\mathrm{rnd}}(k, \Delta)$ begins increasing later as $k$ increases and decays more rapidly as $\Delta$ increases. The TRC coefficients in the random network reaching a local maximum suggest that the club represents a steady-state core structure, thanks to the null model's preservation of permanently stable links. For the club $\mathscr{G}{>280}$, a further increase in $k$ would eliminate the central nodes, Spain and Germany, reducing stability within the club. Therefore, we consider $\mathscr{G}{>280}$ as the most central structure of the N-nutrient trade system. For this, we provide the mathematical definition equation of the stable core club $\mathscr{G}{>k^{*}}$, which must satisfy
\begin{equation}
    M_{\mathrm{rnd}}(k^{*}, \Delta) \geq M_{\mathrm{rnd}}(x, \Delta), \forall \Delta, \forall x \in [0,k^{*}+1].
    \label{Eq:TRC:core:define}
\end{equation}

Combining $M(k, \Delta)$ and $M_{\mathrm{rnd}}(k, \Delta)$, we present the ratio $\mu(k, \Delta)$ results in Fig.\ref{Fig:TRC:M}(d). Utilizing bootstrap statistical tests (Eq.(\ref{Eq:TRC:trendp})), we identify the fraction of statistical significance above 1\% as not having a significant TRC phenomenon, depicted as blanks. The values of $\mu(k, \Delta)$ are shown for the fraction of statistical significance below 1\%. Four typical statistical results are reported in Fig.~\ref{Fig:TRC:M}(e), with the first three being statistically significant. As $k$ increases, the club exhibits a broader distribution. The fourth subplot shows a non-significant result for the core club $\mathscr{G}_{>280}$ in a short-time simulated situation. Recognizing whether it is a true TRC requires a certain time of evolution due to the close links within the club. From the overall results, peripheral nodes (those with $k<150$) in the system show a more pronounced simultaneity in trade relationships. Small economies, influenced by globalization, are integrating into the international trade system, leading to apparent simultaneity. Large economies, with sustained high interconnectivity, are less affected by globalization. However, when the core group $\mathscr{G}_{>280}$ is disrupted at $k>290$, a notable simultaneity effect is observed. The remaining club members no longer maintain permanent relations, and the time trend induces strong simultaneity, reflected in relatively high values of $\mu(290, 24-29)$.

In conclusion, our analysis confirms the presence of the TRC phenomenon in the N-nutrient trade network, utilizing the aggregate degree as the richness metric. Through a comparison between actual TRC coefficients and those simulated by the null model, we identify a central cornerstone club in the network. The collapse of this club marks a shift from a stable to an unstable state. Smaller players are notably influenced by the globalization trend, developing simultaneous trade relations. Given the contemporary rise of trade protectionism and counter-globalization, the phenomenon of synchronization warrants careful consideration.

\subsection{Comparison between the TRC of different richness}
\label{S:result:B}

\begin{figure*}[htp!]
    \centering
    \includegraphics[width=0.995\linewidth]{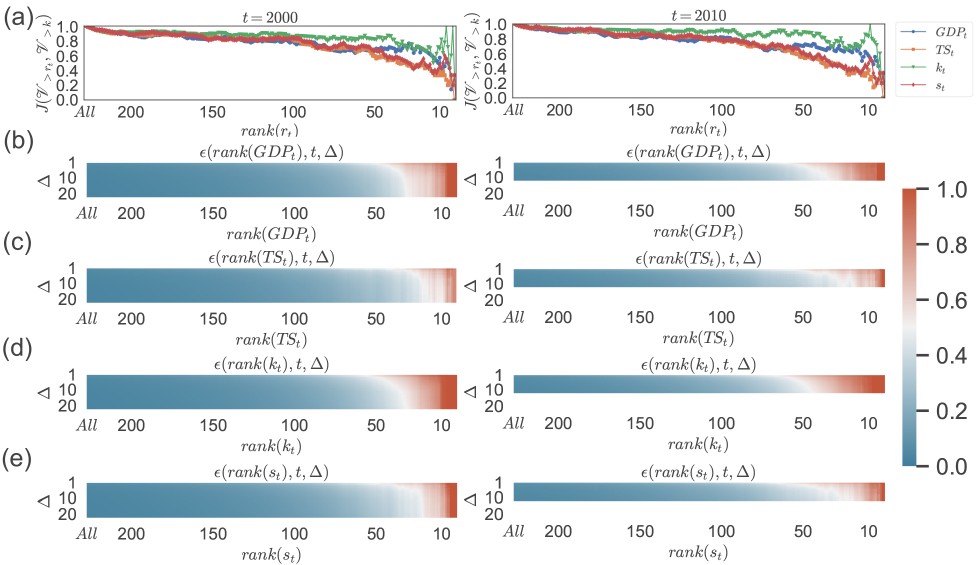}
    \caption{(a) The Jaccard similarity $J(\mathscr{V}_{>r_{t}},\mathscr{V}_{>k})$ between the club sorted by richness $r_t$ and the same size club sorted by the aggregate degree $k$. (b-e) Comparison of the localized TRC coefficients $\epsilon(r_t, t, \Delta$) for the four richness at $t=2000$ and $t=2010$. All four richness are compared in ranked form. (b) Ranked by the $GDP$ of the economy in year $t$. (c) Ranked by the economy's total supply $TS$ in year $t$. (d) Ranked by the economy's instantaneous degree $k$ in year $t$, where $k$ is the sum of out-degree and in-degree. (e) Ranked by the total trade volume $s$ of the economy in year $t$.}
    \label{Fig:TRC:comparsion}
\end{figure*}

In the previous section, we comprehensively examined the overall temporal network from 1991 to 2020 for TRC analysis, using the aggregate degree $k$ as the richness to describe synchronization and clustering phenomena in the system.
However, understanding the formation mechanism of TRC requires extracting more information from the system. We aim to disassemble the overall temporal network layer by layer and observe localized TRC with each initial year $t$, calculated using Eq.~(\ref{Eq:TRC:phi}). Additionally, to investigate the TRC formation mechanism, we need to identify the most critical richness of the nodes for the appearance of TRC. Therefore, we will choose four importance metrics as richness for constructing different localized TRCs for comparison.

Benefiting from the maturity of static network structure research, various properties can be used as richness. Centrality metrics of nodes such as degree, eigenvector, PageRank, betweenness, and closeness can be applied. Weight is a necessary consideration in complex systems \citep{Barrat-Barthelemy-PastorSatorras-Vespignani-2004-ProcNatlAcadSciUSA}, and weighted network variables such as strength and average strength can be used \citep{Opsahl-Colizza-Panzarasa-Ramasco-2008-PhysRevLett, Serrano-2008-PhysRevE}. Variables outside the network can also be used, such as the gross domestic product $GDP_t$. Rich economies are considered to be the most influential nodes, and they tend to occupy the most central part of trade, forming dominant roles. Specifically, for the influence of the economy on the N-nutrient system, it is proportional to the total amount of N-nutrient that an economy can control, which is the sum of the production and import of the economy. Here, we use the total supply $TS$ as a variable, recorded as:
\begin{equation}
    TS_{it} = s^{\mathrm{in}}_{it} + produce_{it},
    \label{Eq:junhen}
\end{equation}
$s^{\mathrm{in}}_{it}$ is the amount of imports in the economy $i$ in year $t$, and $produce_{it}$ is the amount of production in the economy $i$ in year $t$. The higher the $TS_{it}$, the more N-nutrients the economy can make decisions with, and the more naturally it can have a more important position in the N-nutrient system. From the perspective of the trade network, the total trade volume of the economy $s_{it} = s^{\mathrm{in}}{it} + s^{\mathrm{out}}{it}$ is regarded as the main network centrality in the weighted network and the most dominant measure of influence in the actual trade system. Definitely, the most classic centrality indicator, the instantaneous degree $k_t$ in $G(t)$ is a must, calculated as $k_{it} = k_{it}^{\mathrm{out}} + k_{it}^{\mathrm{in}}=\sum_j \left(e_{ijt} + e_{jit}\right)$. All these indicators are collected in the initial year $t$ to analyze the evolution from $t$ to $t+\Delta-1$ as Eq.~(\ref{Eq:TRC:phi}). Summing up the above, we will use four indicators as the richness: the economy's $GDP_t$ as the importance as a whole, the total supply $TS_t$ as the importance in the N-nutrients system, the instantaneous degree $k_t$ as the importance in the unweighted trade network, and the total trade volume $s_t$ as the importance in the weighted trade network. Through these clubs, we intend to compare the difference and clarify the reasons affecting the TRC phenomenon.

In Fig.~\ref{Fig:TRC:comparsion}, we present the results for two initial years, $t=2000$ and $t=2010$, representing different stages of the system. To facilitate comparison between the four different indicators, we standardized the values and ranked them. This is a common approach in rich club studies \citep{Jiang-Zhou-2008-NewJPhys,Zhou-Mondragon-2004-IEEECommunLett}. In Fig.~\ref{Fig:TRC:comparsion}(a), we report the Jaccard similarity $J(\mathscr{V}_{>r{t}},\mathscr{V}_{>k})$ between the rich club of nodes with richness $r_t$ larger than a specific rank and the club of nodes with the aggregate degree $k$ larger than the same specific rank, calculated as: $J(\mathscr{V}_{>r_{t}},\mathscr{V}_{>k})=\frac{\sharp \vert\mathscr{V}_{>r_{t}}\cap \mathscr{V}_{>k}\vert} {\sharp \vert\mathscr{V}_{>r_{t}}\cup \mathscr{V}_{>k}\vert}$. Despite the difference in order, the top-ranked nodes sorted by $k_t$ and those sorted by $k$ tend to be the same economies. This invariance of the instantaneous club to the overall club is one of the keys to the TRC.

After analyzing the results in Fig.~\ref{Fig:TRC:comparsion}(b-e) and the results for other initial years, we observe a consistent pattern:
\begin{equation}
\resizebox{1\linewidth}{!}{
$\begin{aligned}
   & \sum_{rank(k_t),\Delta} \epsilon \left(rank(k_t), t, \Delta \right) > \sum_{rank(GDP_t),\Delta} \epsilon \left(rank(GDP_t), t, \Delta \right) \\ & >  \sum_{rank(s_t),\Delta} \epsilon \left(rank(s_t), t, \Delta \right) > \sum_{rank(TS_t),\Delta} \epsilon \left(rank(TS_t), t, \Delta \right).
    \label{Eq:richness_compoarion}
\end{aligned}$}
\end{equation}
This empirical result suggests that, in the context of localized TRC formation, the order of importance is as follows: instantaneous degree ($k_t$) $>$ GDP ($GDP_t$) $>$ total trade volume ($s_t$) $> $ total supply ($TS_t$). This order aligns with the ranking order of the Jaccard similarity between the rich club based on different indicators and the club based on aggregate degrees $k$. It emphasizes the importance of instantaneous degree ($k_t$) in the study of TRC and highlights the requirements for the formation of a stable rich club. Nodes with high instantaneous degrees ($k_t$) are more prone to forming stable and closely related clubs. These nodes are considered well-known network hubs, leading to increased probability and stability of connections. This effect exists in wealthy economies with less stability. When rich clubs are constructed based on trade volume ($s_t$), the localized TRC phenomenon weakens significantly. Major economies, sorted by the sum of imports and exports, are unlikely to generate localized TRCs, indicating an unstable trade group. The reason could be that major economies often rely more on their own domestic production, as observed in the case of the United States, a significant net importer that still produces about 70\% of its N-nutrients domestically. Although we use the sum of imports and production ($TS_t$) as a proxy, major economies sorted by $TS_t$ are unlikely to generate localized TRCs. For example, India, despite being a global leader in N-nutrient production, neither occupies a central position in the trade network nor establishes a stable import-export structure. This might explain why India has become increasingly dependent on imports instead of self-production from 2000 to 2020. These observations suggest that trade volume or availability is not the most dominant factor in the formation of a stable rich club. In general, the clubs constructed using different variables exhibit different performances, with degree being the most significant factor in generating localized TRCs. This implies that a good reputation and a central position in the trade network play crucial roles in the formation of localized TRCs.

\subsection{Evolution of the TRC of different richness}
\label{S:result:C}

\begin{figure*}[htp!!!!!!]
    \centering
    \includegraphics[width=0.995\linewidth]{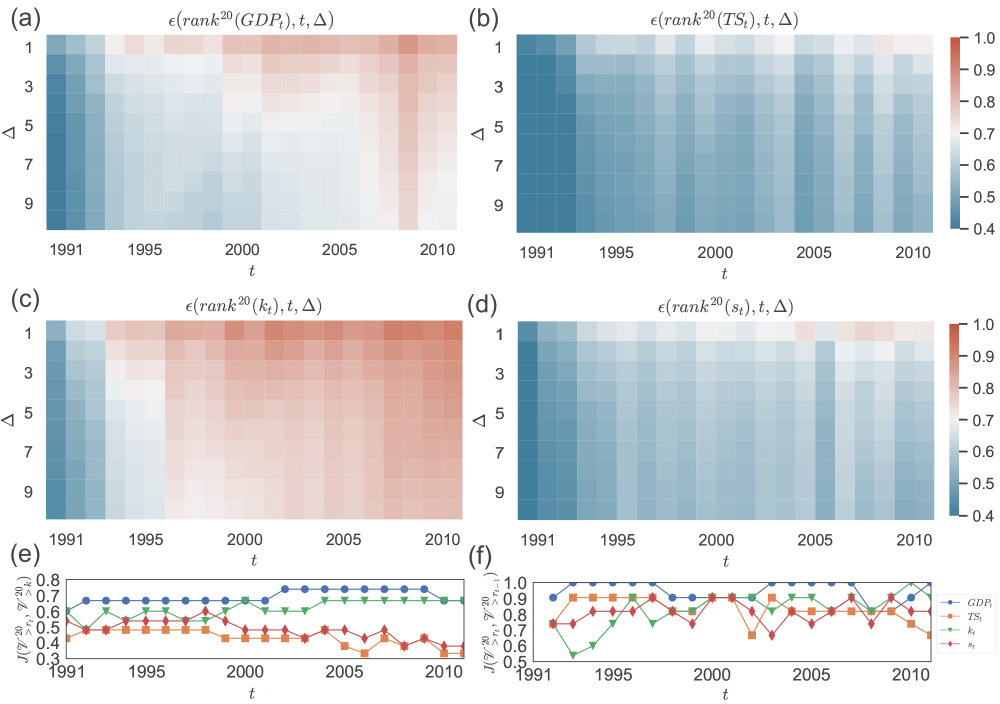}
    \caption{Evolution of the localized TRC coefficients for a fixed rich club of the top 20. (a) The top 20 economies with the highest $GDP_t$. (b) The top 20 economies with the highest $TS_t$. (c) The top 20 economies with the highest $k_t$. (d) The top 20 economies with the highest $s_t$. (e) Evolution of the Jaccard similarity $J(\mathscr{V}^{20}_{>r_{t}},\mathscr{V}^{20}_{>k})$ between the top 20 club sorted by richness $r_t$ and the fixed top 20 club sorted by the aggregate degree $k$. (f) Evolution of the Jaccard similarity $J(\mathscr{V}^{20}_{>r_{t}},\mathscr{V}^{20}_{>r_{t-1}})$ between the top 20 club sorted by richness $r_t$ and that sorted by richness in the last year $r_{t-1}$. }
    \label{Fig:TRC:comparsion:epsilon:evolution}
\end{figure*}

The previous section provides a static view, while comparing evolutionary processes is necessary to explore the transition from localized TRC to TRC. To address this concern, we conducted an evolutionary analysis of the localized TRC coefficients from 1991 to 2011 using the four richness measures. Due to the crucial position of the top-ranking nodes, we use the top 20 as the scale to represent the evolution, recorded as $\mathscr{V}^{20}_{>r_{t}}$. Since $J(\mathscr{V}^{20}_{>GDP_{t}},\mathscr{V}^{20}_{>k})$ and $J(\mathscr{V}^{20}_{>k_{t}},\mathscr{V}^{20}_{>k})$ are similar, the comparability is enhanced. The evolution analysis will provide a clearer understanding of the formation process of TRC in the N-nutrient trade network.

In Fig.~\ref{Fig:TRC:comparsion:epsilon:evolution}, we observe the localized TRC phenomenon for different richness and initial years $t$, using the top 20 economies as the club. The maximum evolution time is set to $\Delta = 10$. Overall, the results of all richness show increasing TRC coefficients over time, reflecting the increasingly dense trade relationships between economies brought about by globalization. For the top 20 $GDP_t$ club in Fig.~\ref{Fig:TRC:comparsion:epsilon:evolution}(a), the most stable state is formed in 2009, during the global economic crisis. The shock highlights the special characteristics of these economies, which may have played a role in protecting their GDP during the crisis. The $TS_t$ has the lowest TRC coefficient in Fig.~\ref{Fig:TRC:comparsion:epsilon:evolution}(b). For the top 20 economies ranked by degree $k_t$ in Fig.~\ref{Fig:TRC:comparsion:epsilon:evolution}(c), the highest TRC coefficient is observed. From 1991 to 1996, there is a significant increase in both stability and correlation. After that, the club maintains high stability while having a tight correlation. While node strength $s$ is important for judging whether nodes will be related to each other in a weighted network, it is not as significant in the analysis of TRC. In Fig.~\ref{Fig:TRC:comparsion:epsilon:evolution}(d), the localized TRC coefficient for the top 20 clubs ranked by $s_t$ is low, with only some tight relationships observed after 2008 in the static network, and relatively poor stability over time.

Through the comparison of the evolution of the localized TRC coefficient for the four richness, a clear pattern emerges: for the vast majority of $\Delta$ and $t$, it has
\begin{equation}
\begin{aligned}
     & \epsilon(rank^{20}(k_t), t, \Delta) >  \epsilon(rank^{20}(GDP_t), t, \Delta) \\ & >  \epsilon(rank^{20}(s_t), t, \Delta) > \epsilon(rank^{20}(TS_t), t, \Delta).
    \label{Eq:richness_compoarion:evo}
\end{aligned}
\end{equation} 
This pattern confirms that nodes with high instantaneous degrees are tend to form the most stable central structures, a feature that persists throughout the time evolution, indicating that reputation and network importance are the vital factors to TRC.

It is worth noting that $J(\mathscr{V}^{20}_{>GDP_{t}},\mathscr{V}^{20}_{>k})$ is usually higher than $J(\mathscr{V}^{20}_{>k_{t}},\mathscr{V}^{20}_{>k})$ in Fig.~\ref{Fig:TRC:comparsion:epsilon:evolution}(e). $k_t$ captures the club that is most stable and tight in the short term based on the instantaneous structure $G(t)$. If we only calculate the localized TRC at moment $t$, $k_t$ is actually better than $k$. $TS_t$ and $s_t$ are increasingly failing to capture the most stable parts of the system. Because the weights elevate the importance of certain peripheral nodes, this explains the low TRC coefficients. On the other hand, based on Fig.~\ref{Fig:TRC:comparsion:epsilon:evolution}(f), unlike the stable $GDP_t$ ranking, the $k_t$ ranking is changeable, with a trend from unstable to stable. It implies the structural information from $k_t$ needs suitable extrapolation and simulation to ensure correctness in the long term.

Despite the changing club membership, there is an overall trend,
\begin{equation}
     \epsilon(rank^{20}(k_t), t, \Delta) >  \epsilon(rank^{20}(k_t), t-1, \Delta),
    \label{Eq:richness_compoarion:t}
\end{equation}
indicating that the core club maintains an expanding trend over time to ensure its status in the growing system. The key of TRC with low $\Delta$ lies on the latter part of the system, while that with high $\Delta$ lies on the former part. The earlier a strong localized TRC is formed, the stronger the TRC phenomenon can be produced, which still shows the importance of degree. All these suggest that the degree is the most critical richness of nodes to the formation of TRC from the view of year by year.

\subsection{Formation mechanisms of the TRC phenomenon}
\label{S:result:D}

We have identified the TRC phenomenon that exists in the N-nutrient trade network in section A and the primary factor in the formation of TRC is the node's degree in section B and C. To further understand the mechanism behind this phenomenon, we need to construct complex network models to investigate why TRC occurs. Since the degree is the key factor, it is reasonable to build evolution models based on degree. The network evolution model is denoted by $\mathscr{G}_\mathrm{model}$, based on a certain mechanism from the actual trade network. The model network is built on the structure of the actual network before moment $t$ and evolves independently.
When a model generates a similar TRC as the actual system, the mechanism behind the evolution model can be considered a possible answer to why TRC occurs, or at least a mathematical explanation for it.


The TRC is composed of localized TRCs, so simulating the TRC requires building the localized TRC year by year first. In order to simulate the formation of the localized TRC, the model networks need to evolve at a rate close to the actual network.
Accordingly, we introduce two variables from $\mathscr{G}$, one is the survival rate of links from $G(t)$ to $G(t+1)$
\begin{equation}
    p_{\mathrm{\mathrm{sur}}}(t)= \frac{\sharp\left[{E(t)\cap E(t+1)}\right]}{\sharp\left[{E(t)}\right]}.
    \label{Eq:T}
\end{equation}
It means the average probability that a link in $G(t)$ reoccurs in $G(t+1)$.
Another is the birth rate of links from $G(t)$ to $G(t+1)$,
\begin{equation}
    p_{\mathrm{birth}}(t)= \frac{\sharp\left[{E(t+1)}\right]-\sharp\left[{E(t)\cap E(t+1)}\right]}{\sharp \left[\mathscr{V}_{> r}\right](\sharp \left[\mathscr{V}_{> r}\right]-1)-\sharp\left[{E(t)}\right]}.
    \label{Eq:TRC:p_birth}
\end{equation}
It means the average probability that a link not in $G(t)$ appears in $G(t+1)$. Here, we can directly construct the base model $\mathscr{G}_{\mathrm{er}}$ that breaks or generates links with equal probability, similar to the Erd{\H{o}}s–R{\'{e}}nyi (ER) model. The set of nodes is kept constant $\mathscr{V}$, and the set of links is updated as $E_{\mathrm{er}}(t)=\left\{e^{\mathrm{er}}_{ijt}\right\}$, while the evolutionary process of all links is 
\begin{equation}
    e^{\mathrm{er}}_{ij(t+1)} = e^{\mathrm{er}}_{ijt} h \left(  \xi, p_{\mathrm{\mathrm{sur}}}(t) \right)   + (1 - e^{\mathrm{er}}_{ijt}) h \left( \xi, p_{\mathrm{birth}}(t) \right).
    \label{Eq:TRC:er_evo}
\end{equation}
Meanwhile,
\begin{equation}
    h(a,b)= \begin{cases}1, \ &if \ a \leq b,\\
    0, \ &if \ a > b.
    \end{cases}
    \label{Eq:TRC:h}
\end{equation}
$\xi$ is a random number uniformly distributed in $[0,1]$. The value of $\xi$ is reassigned independently each time. In this way, multiple simulations can be performed to make sure the reliability of the results.

Highly developed economies are more likely to continue to have trade relations with each other \citep{Hallak-2010-RevEconStat}, and nodes with high out-degrees are more likely to connect to nodes with high in-degrees \citep{Barabasi-Jeong-Neda-Ravasz-Schubert-Vicsek-2002-PhysicaA}, which is widely known as preferential attachment (PA). The mathematical hypothesis given for this is that the probability of trade occurring from economies $i$ to $j$ is proportional to the product of their out-degree and in-degree, i.e., $\Pi(i,j) \sim k^{\mathrm{out}}_i k^{\mathrm{in}}_j$. Based on this, we constructed the network model $\mathscr{G}_{\mathrm{pa}}$. $p_{\mathrm{\mathrm{sur}}}(t)$ and $p_{\mathrm{birth}}(t)$ are invoked into controlling the density of generated network. The link evolution process can be inscribed as
\begin{equation}
    e^{\mathrm{pa}}_{ij(t+1)} =  e^{\mathrm{pa}}_{ijt} h \left ( \xi,   \widehat{\beta}^s_t k_{it}^{\mathrm{out}} k_{jt}^{\mathrm{in}}   \right)  + (1-e^{\mathrm{pa}}_{ijt}) h \left( \xi,  \widehat{\beta}^b_t k_{it}^{\mathrm{out}} k_{jt}^{\mathrm{in}} \right).
    \label{Eq:TRC:kk_evo}
\end{equation}
$k_{it}^{\mathrm{out}}$ and $k_{jt}^{\mathrm{in}}$ respectively represent the instantaneous out-degree of node $i$ and in-degree of node $j$ in $G_{\mathrm{pa}}(t)$. The parameters are
\begin{equation}
    \widehat{\beta}_t^{s} = \frac{p_{\mathrm{\mathrm{sur}}}(t)\sharp\left[{E_{\mathrm{pa}}(t)}\right]}{\sum_{E_{\mathrm{pa}}(t)}k_{it}^{\mathrm{out}} k_{jt}^{\mathrm{in}}},
\end{equation}
and
\begin{equation}
    \widehat{\beta}_t^{b} = \frac{p_{\mathrm{birth}}(t)\sharp\left[{\overline{E_{\mathrm{pa}}}(t)}\right]}{\sum_{\overline{E_{\mathrm{pa}}}(t)}k_{it}^{\mathrm{out}} k_{jt}^{\mathrm{in}}}.
\end{equation}
$E_{\mathrm{pa}}(t)$ is the set of $e^{\mathrm{pa}}_{ijt}=1$, and $\overline{E_{\mathrm{pa}}}(t)$ actually refer to the set of $e^{\mathrm{pa}}_{ijt, i\neq j}=0$.

\begin{figure*}[ht]
    \centering
    \includegraphics[width=0.95\linewidth]{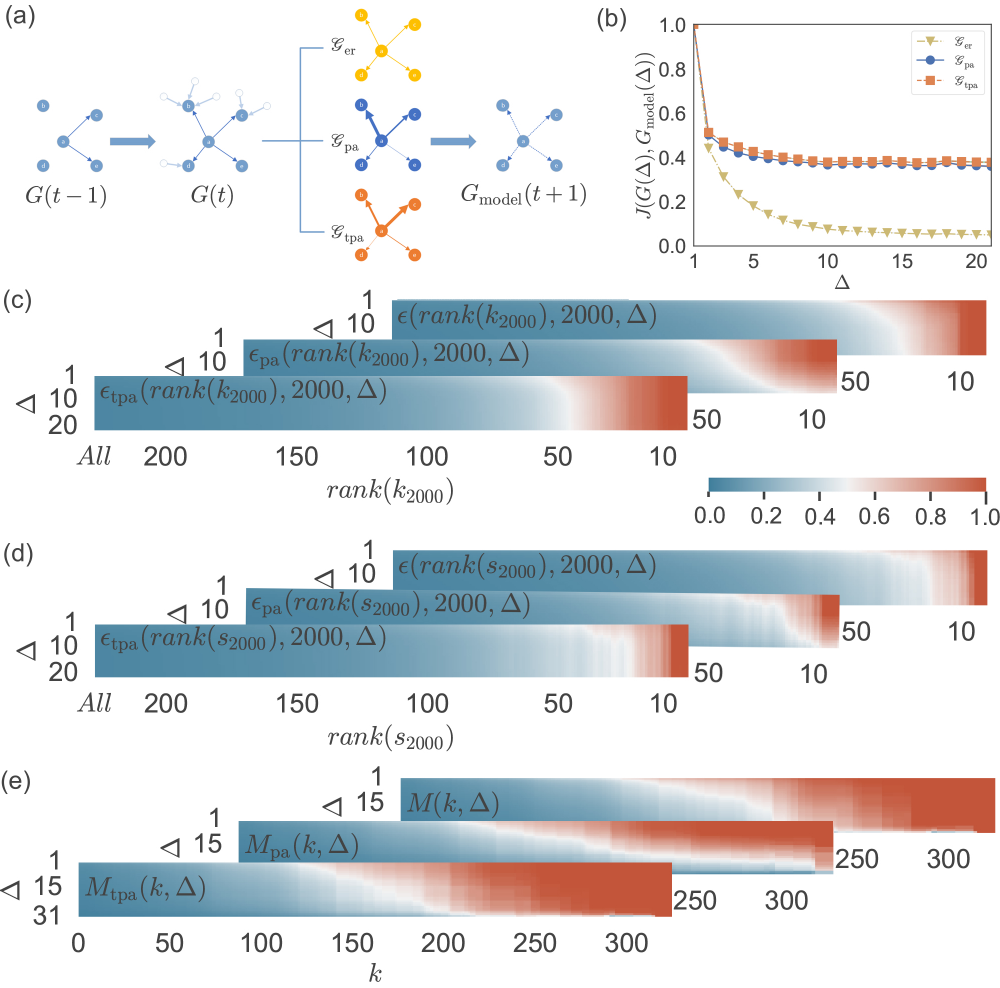}
    \caption{(a) A simple schematic diagram to demonstrate the difference between the three models in the probability of links occurrence, which is proportional to the width of lines. (b) The Jaccard Index of the actual network versus the three model networks, namely, Erd{\H{o}}s–R{\'{e}}nyi (ER) models, the preferential attachment (PA), and the temporal preferential attachment (TPA), revealing the extent to which each model captures the features of the actual network. (c) Comparison of the localized TRC coefficients ranked by the instantaneous degree $k_{2000}$ for the actual network, the PA model, and the TPA model. (d) Comparison of the localized TRC coefficients ranked by the instantaneous strength $s_{2000}$ for the actual network, the PA model, and the TPA model. (e) Comparison of the TRC coefficients $M(k,\Delta)$ ranked by the aggregate degree $k$ for the actual network, the PA model, and the TPA model.
    }
    \label{Fig:TRC:tdpa:comparsion}
\end{figure*}

However, the result reveals that $\mathscr{G}_{\mathrm{pa}}$ only captures a partial picture of the TRC phenomenon. It cannot create a totally stable core club similar to that observed in the N-nutrient trade system. For further exploration, based on the fact that longer-lasting links in the N-nutrient trade are more likely to persist \citep{Li-Wang-Xie-Zhou-2023-JStatMech}, we introduce another formation mechanism, referred to as a kind of path-dependence. If economy $i$'s past decision was to import N-nutrients from economy $j$, then the probability that $i$ imports from $j$ now rises. Here, the trade disruption probability of $e_{ijt}$ is inversely proportional to its existing duration $\tau_{ijt}$ in the history. The mathematical hypothesis is $1 - \Pi(i,j) \sim \frac{1}{\tau_{ijt}}$. We name this evolution model that incorporates this mechanism the temporal-preferential attachment (TPA) model, whose link evolution process is described as:
\begin{equation}
\begin{aligned}
    e^{\mathrm{tpa}}_{ij(t+1)} = & e^{\mathrm{tpa}}_{ijt} h\left (\xi, \widehat{\beta}^s_t (1 -  \frac{\beta^{\tau}}{\tau_{ijt}})  k_{it}^{\mathrm{out}} k_{jt}^{\mathrm{in}} \right)  + \\ & (1-e^{\mathrm{tpa}}_{ijt}) h \left( \xi, \widehat{\beta}^b_t k_{it}^{\mathrm{out}} k_{jt}^{\mathrm{in}} \right),
    \label{Eq:TRC:tkk_evo}
\end{aligned}
\end{equation}
while
\begin{equation}
    \widehat{\beta}_t^{s} =  \frac{ p_{\mathrm{sur}}(t) \sharp\left[E_{\mathrm{tpa}}(t)\right]}{\sum_{E_{\mathrm{tpa}}(t)} \frac{(\tau_{ijt}-\beta^{\tau})k_{it}^{\mathrm{out}} k_{jt}^{\mathrm{in}}}{\tau_{ijt}}},
\end{equation}
and
\begin{equation}
    \widehat{\beta}_t^{b} = \frac{p_{\mathrm{birth}}(t)\sharp\left[{\overline{E_{\mathrm{tpa}}(t)}}\right]}{\sum_{\overline{E_{\mathrm{tpa}}(t)}}k_{it}^{\mathrm{out}} k_{jt}^{\mathrm{in}}}.
\end{equation}
$\tau_{ijt}$ is the existing duration of the link $e_{ijt}$. $\tau_{ijt}=x$ means $e_{ij}$ exists from $G_{\mathrm{tpa}}(t+1-x)$ to $G_{\mathrm{tpa}}(t)$. $\beta^{\tau}$ is estimated from actual data, and here $\beta^{\tau} \approx \frac{1}{2}$. 

We generated the base model, the ER model, and two mechanistic models, the PA model and the TPA model. A demo of the differences between models is represented in Fig.~\ref{Fig:TRC:tdpa:comparsion}(a). For node $a$, the probability of link occurrence with four other nodes $bcde$ is equal in $\mathscr{G}_{\mathrm{er}}$. In $\mathscr{G}_{\mathrm{pa}}$, the probability is proportional to another node's $k^{\mathrm{in}}$, since node $a$'s $k^{\mathrm{out}}$ is the same. The probability of $e_{ac}$ and $e_{ae}$ will rise in $\mathscr{G}_{\mathrm{tpa}}$ because of their presence in $G(t-1)$. The set including $\{G(t), G_{\mathrm{model}}(t+1), G_{\mathrm{model}}(t+2)....G_{\mathrm{model}}(t+\Delta-1) \}$ will be used to calculate the localized TRC coefficient $\epsilon_{\mathrm{model}}(r, t, \Delta)$. The TRC coefficient $M_{\mathrm{model}}(r, \Delta)$ can then be obtained by iterating over all $\epsilon_{\mathrm{model}}(r, t, \Delta)$.

In Fig.~\ref{Fig:TRC:tdpa:comparsion}(b), we compare these three simulated networks with the actual network at the same evolution time $\Delta$ using the Jaccard index \citep{Palla-Barabasi-Vicsek-2007-Nature}, calculated as $J(\mathscr{G}, \mathscr{G}_{\mathrm{model}})=\frac{\vert{\mathscr{E}}\cap {\mathscr{E}_{\mathrm{model}}}\vert} {\vert{\mathscr{E}}\cup {\mathscr{E}}_{\mathrm{model}}\vert}$. We set the initial year to 2000. Results show that $\mathscr{G}_{\mathrm{pa}}$ and $\mathscr{G}_{\mathrm{tpa}}$ preserved a similar proportion of links to the actual trade network as they evolved. After an initial decrease in similarity, both models maintain a stable portion of about 0.4 of the actual network. Note that the number of the same links has practically increased over time as the network density increases. Comparing the two models, $\mathscr{G}_{\mathrm{tpa}}$ shows a slight advantage in maintaining similarity. In contrast, the similarity of $\mathscr{G}_{\mathrm{er}}$ with the actual network continuously decreased over time, approaching that of a completely independent distribution. Overall, the PA and TPA models preserve some properties of the actual network through the evolutionary process, allowing us to analyze whether their mechanisms contribute to the formation of the TRC phenomenon.

As models based on instantaneous degree evolution, the first thing to consider is the localized TRC coefficient when the nodes' $k_t$ is the richness.
In Fig.~\ref{Fig:TRC:tdpa:comparsion}(c), we show the results when the initial year is set to the year $t=2000$ as representative. The right graph shows the localized TRC coefficient in the real system. The middle graph depicts the localized TRC result of $\mathscr{G}_{\mathrm{pa}}$. We can observe a TRC phenomenon, but weaker than the actual situation. The core experiences decay, and the system tends towards decentralization. The result of $\mathscr{G}_{\mathrm{tpa}}$ is shown in the left graph. It is much closer to the actual situation and retains strong stability in the core structure over long periods of evolution. 

Then we analyze the performance of the other rich clubs. In Fig.~\ref{Fig:TRC:tdpa:comparsion}(d), the results sorted by the strength $s_t$ are reported. Though the calculations of the PA and TPA models are unrelated to $GDP_t$, $TS_t$, and $s_t$, all these rich clubs of $\mathscr{G}_{\mathrm{tpa}}$ show high similarity to the actual situation. Stable clubs triggered by degrees naturally extend into other clubs that have some correlation with degrees. It can be a reasonable explanation for the observed TRC phenomenon of other variables in the actual system.

In the end, we iterate $\epsilon_{\mathrm{model}}(k, t, \Delta)$ at each initial time $t$ to calculate the TRC coefficient $M_{\mathrm{model}}(k, \Delta)$ to compare with $M(k, \Delta)$. The collection is similarly calculated as $M_{\mathrm{model}}(k,\Delta) =\max_{t} \epsilon_{\mathrm{model}}(k, t, \Delta)$, where $k$ is the aggregate degree. The comparison is in Fig.~\ref{Fig:TRC:tdpa:comparsion}(e). A distinct difference between $M_{\mathrm{pa}}(k,\Delta)$ and $M(k, \Delta)$ lies on whether the core club is stable or not. The PA mechanism alone is not sufficient to explain the formation of the strong TRC phenomenon. On the other hand, $M_{\mathrm{tpa}}(k,\Delta)$ may be stronger than $M(k, \Delta)$. Because the model cannot capture the impact of exogenous events, such as the 2008 economic crisis and the COVID-19 pandemic, which can cause significant disruptions and lead to the collapse of stable rich clubs. The model is ideal. We have also constructed a model with only the path-dependence mechanism and found that the results are completely off the mark. Therefore, we have the conclusion that the TRC phenomenon in the N-nutrient trade network depends on the combination of degree-homophily and path-dependence and cannot be generated without either one.

\section{Discussion and summary}
\label{S:disscusion}
 
To summarize, we have developed an effective network model to simulate the temporal rich club phenomenon in the N-nutrient trade network, which is crucial for global food security and ecological security. The model helps us understand the formation mechanism of TRC. Before modeling, our preparations included identifying the statistically significant TRC phenomenon in the N-nutrient trade network. The most stable cornerstone group in the N-nutrient trade network from 1991 to 2020 includes the Netherlands, the United States, France, China, the United Kingdom, Belgium, Germany, and Spain. From this group, other economies form a comprehensive trade network. Due to the ongoing trend of globalization, both unstable components and smaller economies exhibit temporal simultaneity.

After analyzing the localized TRC of different initial years and richness, our analysis showed that degree is the most dominant factor for TRC. It is a prerequisite for modeling evolution, which is based on the nodes' degree. The rich club, sorted by instantaneous degree, has the strongest static correlation and temporal stability, followed by GDP, trade volume, and total supply. It indicates that building upstream and downstream trading partners to establish reputation and influence is more important than initially thought. The association structure between well-connected nodes becomes tighter and more stable over time. Additionally, the oligopoly alliance tends to maintain an expanding trend.


Regarding the formation of TRC, we incorporated two mechanisms into the mathematical evolution model. The first mechanism suggests that nodes with high export degrees tend to export to nodes with high import degrees, characterized as degree-homophily driven by supply-demand and reputation attraction. It is related to the preferential attachment in complex networks and trade gravity in economics. The second mechanism proposes that the longer the trade lasts, the more stable it becomes, akin to path-dependence explained by communication costs, comparative advantages, and scale effects. Through our model, we confirmed that the simultaneous existence of these two mechanisms leads to the emergence of a TRC similar to the actual system during the theoretical model's evolution. The coexistence of these mechanisms results in an oligopoly alliance controlling the entire system over long-term evolution.

While our attempt to explain the formation of TRC in the N-nutrients trade network provides a novel perspective on the evolution of the system and the prominent nodes set, it is essential to note the limitation of our study being focused on a single system. The conclusions and models may not be directly applicable to other networks, given that the intrinsic mechanisms of different systems can vary. Since our mechanism-based evolutionary models are effective approach for understanding the formation of TRC, our work can still serve as a reference for the in-depth exploration of the TRC phenomenon in various networks.


\section*{Acknowledgments}

This work was partly supported by the National Natural Science Foundation of China (72171083) and the Fundamental Research Funds for the Central Universities.


%

\end{document}